\documentclass[%
 reprint,
 amsmath,amssymb,
 aps,
]{revtex4-1}

\usepackage{graphicx}% Include figure files
\usepackage{dcolumn}% Align table columns on decimal point
\usepackage{bm}% bold math
\usepackage{color}
\newcommand{\tI}{t_i}
\newcommand{\tF}{t_f}
\newcommand{\fun}{{\cal F}}
\newcommand{\syst}{{\bf S}}
\newcommand{\env}{{\bf E}}
\newcommand{\Gh}{\Gamma}
\newcommand{\GhI}{\Gh_i}
\newcommand{\GhF}{\Gh_f}
\newcommand{\GA}{\Gh_\syst}
\newcommand{\G}{G}
\newcommand{\GsI}{\G_i}
\newcommand{\GsF}{\G_f}
\newcommand{\Om}{\Omega}
\newcommand{\om}{\Omega}
\newcommand{\Gs}{\G}
\newcommand{\SI}{{\cal S}}
\newcommand{\dis}{\Sigma}
\newcommand{\del}{\gamma}
\newcommand{\Q}{\Psi}
\newcommand{\QQ}{\psi}
\newcommand{\F}{\Phi}
\newcommand{\xx}{x}
\newcommand{\xxI}{\xx_i}
\newcommand{\xxF}{\xx_f}
\newcommand{\st}{s}
\newcommand{\stI}{\st_i}
\newcommand{\stF}{\st_f}
\newcommand{\xa}{z}
\newcommand{\xaI}{\xa_i}
\newcommand{\xaF}{\xa_f}
\newcommand{\yw}{\alpha}
\newcommand{\ywI}{\yw_i}
\newcommand{\ywF}{\yw_f}
\newcommand{\OS}{{\cal O}}
\newcommand{\info}{{\cal I}}
\newcommand{\infJ}{{\cal J}}
\newcommand{\pin}{P_{in}}
\newcommand{\pout}{P_{out}}
\newcommand{\pr}{p}

\newcommand{\ha}{{\cal H}}
\newcommand{\W}{W}
\newcommand{\Wen}{W_{tr}}
\newcommand{\Qh}{Q}
\newcommand{\E}{\epsilon}
\newcommand{\DE}{\Delta\E}

\newcommand{\e}{e}

\begin{document}

\preprint{APS/123-QED}

\title{Information form of the second law of thermodynamics
 }% Force line breaks with \\

\author{Miroslav Hole\v{c}ek}
 \email{holecek@rek.zcu.cz}
\affiliation{%
 New Technologies Research Center, University of West Bohemia, Plze\v{n} 301 00,  Czech Republic
}%

\date{\today}% It is always \today, today,

\begin{abstract}

An essential role of information in microscopic thermodynamics  (e.g. Maxwell's demon) opens a challenging question if there exists a formulation of the second law of thermodynamics based  only on  pure information ideas.  Here,  such a formulation is suggested for unitary processes by introducing information  as a full-valuable physical quantity defining an (objective) microscopic information entropy as 'information about microstate'. We show that various forms of entropy (Boltzmann, Shannon, Clausius) are in fact only a special cases of information entropy whose general form is found out.  An observer plays here the role of a special (information) reference frame (IRF) towards which the entropy is defined.  Some paradoxes or misunderstandings connected with the concept  of entropy or the content of the second law arise by describing a situation without specifying a concrete  IRF. Typically, the Boltzmann statistical approach cannot be symmetrically used towards the past as towards the future in one IRF. The information second law is full usable at meso- or microscopic scales: the information form of the generalized second law is found out too.

%\begin{description}
%\end{description}
\end{abstract}

% \pacs{Valid PACS appear here}% PACS, the Physics and Astronomy
\maketitle

%\tableofcontents

\section{Introduction}
 As noticed by J.C. Maxwell in 1871, if a being ("demon") has \emph{information} about individual molecules of a system and is able to manipulate with them, the second law of thermodynamics can be violated \cite{Maxwell1871}. The patch to this "hole in law" has been found in the last decades: demon's information about molecules has to be taken as a full-valued physical  component  of the description of the whole situation \cite{MarNorVed,SagUed2012,ParHorSag,DefJar2013,BarSei2014b,ShiMatSag2016}. This idea has been demonstrated and verified in various experiments at meso- and nanoscales \cite{Toy2010,Mih2016,Cot2017}, it has surprising
applications in molecular biology \cite{BaSei2013,HorSagPar2013} and nanoscale technologies \cite{Seifert2012,TanShe2008}. A general physical theory of information addressing the basic problems of statistical physics (macroscopic time-asymmetry, objective microscopic definition  of entropy) \cite{ParHorSag}, however, does not exist yet. 

In such a theory, information should be a  basic concept expressing someone's knowledge about a physical system \cite{mutual}. To bring this idea closer, consider a system observed by several observers  so that each of them has  a different knowledge about it.
 The perfect observer \cite{Laplace1812} knows its microscopic state,  $\xx$, others know less. The knowledge of individual observers can be quantified  by using the ingenious idea by C.E.~Shannon \cite{Shannon1948}. Imagine that the observers receive the message fully describing  $\xx$.   
The \emph{value of information} included in this message is different for each of them  (for the perfect observer the message is worthless). Denote as $\info_\OS (\xx )$ this value for an observer $\OS$. We can say that this value quantifies observers' ignorance (lack of information) as to $\xx$ before receiving the message.  

If $\OS =\OS (M)$ is a typical macroscopic observer then $S=\info_{\OS (M)} (\xx )$ is the definition of  entropy by E.T.~Jaynes \cite{Jaynes1957}. Entropy is then a function of microscopic state $\xx$. Its evident dependence on a "typical macroscopic observer" is, however, incompatible with objective physics. But notice that the observer in the definition of $\info_\OS (\xx )$ plays only the role of some informational \emph{reference} with respect to which information in the message is assessed.     
The fact that the value of a quantity is defined only with respect to another value of this quantity is not unfamiliar in physics.

To determine the position of a body, for example, we must do it in reference to positions of other bodies.
The body  has undoubtedly an objective location in space, its concrete meaning, however, is given in a concrete reference frame.  The concept of  microscopic information entropy, $S=\info_{\OS (M)} (\xx )$, is analogical. It can be understood as an abstract objective quantity expressing  information  connected with a given microscopic state $\xx$. Its concrete value and meaning, however, must be specified in a concrete \emph{information reference frame} (IRF) represented, for example, by a macroscopic observer $\OS (M)$. 

We do not observe  processes  at the very  microscopic level and  information about them is usualy given indirectly by observing other (usualy some coarse-grained) quantities, which is the original idea  by L.~Botzmann \cite{Boltzmann1896}. Information gained by observing a system at an arbitrary spatial scale  is what  define a concrete IRF. It implies that the role of a macroscopic observer is not exclusive. We can have arbitrary information reference frames defined with respect to our experimental abilities or our intentions to model a concrete system. Information entropy - an objective quantity - has different values at these reference frames.  A full microscopic description of a system (e.g. as a pure quantum state) is also a special (microscopic) information frame. 
 
In this contribution, we use the idea of information reference frames  to derive a novel formulation of  the second law of thermodynamics 
which introduces information as a fundamental physical quantity without a need to define a particular observer and to introduce the concept of probabilities. 
The law is formulated for single deterministic and time-symmetric microscopic trajectories that play the role of adiabatic processes at arbitrary spatial scales. 

The finding of an explicit relation between information and entropy allows us to derive a general form of the entropy dependence on the microscopic state of the system in a concrete IRF, 
$S_\Gh (\xx )= k_B (\ln |\Gh|_\Q +\Q_\Gh (\xx ) )$, where $\Gh$ is a subset of the state space determined by a chosen IRF, $|\Gh|_\Q$ is a special measure,  and $\Q$ is a real function whose concrete choice (the so-called $\Q$-representation) defines a concrete form of the entropy. A special choice of $\Q$ defines probabilities, $p$, and the Shannon entropy, $S=-k_B\ln p$. 

The information form of the second law is in fact an information generalization of Boltzmann's ingenious ideas \cite{Boltzmann1896,Lebowitz1993,GLTZ2019}. The information approach, however, avoids the problematic point of Boltzmann's statistical program: namely the fact that his statistical method gives peculiar results when applied towards the past. We argue that the way of thinking towards the past is not symmetric from information point of view: it must be done in another information reference frame. The paradox then disappears. The use of our approach into thermodynamics defined at microscopic or mesoscopic scales is straightforward. We can connect the change of entropy with the dissipation work. The generalized second law is presented here in a  pure information form.

The paper is organized as follows. The concept of value of information and the information reference frame are introduced and  the information entropy is defined in Section~\ref{sec:IRF}. The general definition of adiabatic processes is formulated and the information formula described the change of information entropy in these processes is found out in Section~\ref{sec:AD}. In Section~\ref{sec:GB}, the concrete form of information entropy (i.e. its dependence on the microstate of the system and the used information reference frame) is derived. It generalizes the Boltzmann entropy. The information form of the second law of thermodynamics is discussed in Section~\ref{sec:SLT}. 
   Special forms of the information entropy (the so-called $\Q$-representations) are found out in Section~\ref{sec:SC}: the Shannon entropy and Clausius entropy of classical equilibrium thermodynamics. In Section~\ref{sec:prob}, the concept of probability arising in a special $\Q$-representation of the information entropy is studied, especially its referential meaning given by a special choice of the information reference frame.  We show in Section~\ref{sec:TD}  that some concepts of nonequilibrium statistical physics and the generalized second law of thermodynamics are special applications of the information second law.

\section{Information reference frame and entropy}
\label{sec:IRF}

Information about a system $\syst$  can be identified with a subset $\om$ of its microscopic state space $\GA$.   
Namely the set  $\om$ can be interpreted as including all microscopic states that are consistent with attainable information about the system observed (studied) by a concrete observer. A typical example is Boltzmann's macrostate that includes all microstates that look the same for a macroscopic observer. A subset of the microscopic state space $\om\subset\GA$ may represent, however, arbitrary  information about the system gained at micro-, meso- or macroscopic scales. 

If we want to quantify information we need some \emph{referential information} that plays the role of an information unit. Consider that an observer knows that the actual microstate belongs into a set $\Gh$. The set $\Gh$ plays the role of referential information. The value of information that the actual microstate, $\xx$, belongs into a set $\om\subset\Gh$ can be defined with respect to referential information "$\xx\in\Gh$"  as follows. 
Let the observer receive the message that "$\xx\in\om$".  
 The \emph{information value} of this message (for this observer) is a real number $\info_\Gh (\om )\ge 0$. If $\om =\Gh$ this value is obviously zero (the message is worthless), i.e. $\info_\Gh (\Gh )=0$. The larger the information value $\info_\Gh (\om )$ the more valuable information  is gained by the observer   after receiving the message "$\xx\in\om$". 
 
It is important to say that  the referential set  $\Gh$ is not an arbitrarily chosen subset of the state space. It plays the role of an 'information content' of a concrete observer, whereas 'observer' may be an experimental device, a robot with sensors, a typical macroscopic observer, etc. If an observer connected with  referential information $\Gh$ gains additional information about the system, say "$\xx\in\om_0$", the referential set changes into $\Gh'=\Gh\cap\om_0$.

A collection of referential sets  $\Gh (t)$ for various $t$ during a single microscopic process  is called the \emph{information reference frame} (IRF) in which this process is studied.
  Let an observer (or observers) study a process in between times $\tI$ and $\tF$. The IRF may be formed by  all $\Gh (t)$, $t\in \langle \tI ,\tF \rangle$, if the referential set can be defined at each time moment (e.g. when the process can be continuously observed). An extreme case of IRF describing  a process from $\tI$ to $\tF$ is the IRF formed only by $\Gh (\tI )$ and $\Gh (\tF )$. 
 
\begin{figure}
\includegraphics[width=0.4\textwidth]{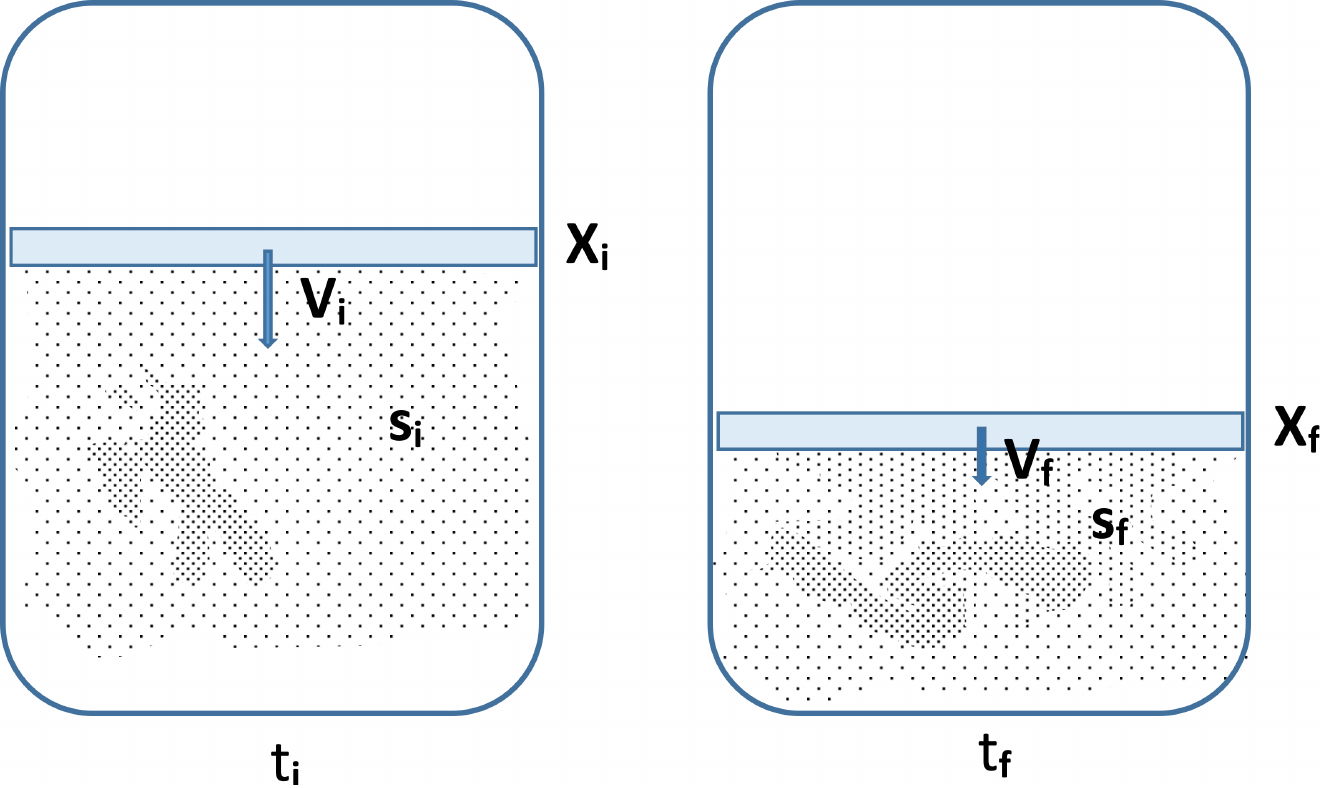}
\caption{ An example of IRF. Consider an isolated system including a horizontal, free-movable massive  piston in a gravity field and a gas occupying the space bellow the piston. The set of observable indicators consists of the position of the piston, $X$, and its current velocity $V$ (in vertical direction), i.e. $\yw =(X,V)$. At $\tI$, the gas is in an arbitrary (non-equilibrium) state $\stI$, the piston at the state $\ywI =(X_i,V_i)$. The piston is  falling down so that it has the position $X_f$ and velocity $V_f$ at $\tF >\tI$ while the gas reaches the  state $\stF$. At $t$, the set $\overline{\Gh} (\yw (t))$  is formed by \emph{all} possible states of the gas being consistent with the current state of the piston, $\yw (t)$. It implies that $\GhI\equiv\Gh (\tI )$ includes many states that \emph{cannot} evolve the system to the state in which the piston state is $\ywF$ at $\tF$, and the set $\GhF\equiv\Gh (\tF )$ includes many states that cannot be reached from the given initial conditions at $\tI$. See Fig.~\ref{fig:2}.  }
\label{fig:1}
\end{figure}

The important situation occurs if referential sets can be identified with values of some physical quantities $\yw$, we call them  the \emph{observable indicators}. The IRF is then defined by time evolution of these quantities (see Fig.~\ref{fig:1}).  
The set $\Gh =\Gh 
 (\yw )$ then includes all microscopic states that are consistent with the value of $\yw$. Typical indicators are various coordinates defining the thermodynamic state \cite{Jaynes1965,Callen1985}   regardless  at which length scale the observation  is done (volume of a gas, quantities at  small hydrodynamic cells, distance of the ends of a macromolecule \cite{Jar2011}).
 
The observable indicators may be fully controlled by an observer (e.g.  driven by an external agent  \cite{Jar1997,Crooks1998,Crooks2000}) or may be understood as special degrees of freedom of a fully autonomous system. In the latter case, the microstate $\xx$ of the system equals $\xx =(\st ,\yw )$, where $\st$ are  \emph{internal} (not-observed) degrees of freedom  (see Fig.~\ref{fig:1}), and 
\begin{equation}
\Gh (\yw ) =\{(\st , \yw ),\,\, \st\in \overline{\Gh} (\yw )\}.
\end{equation} 
The set $\Gh$ may represent a "grain" in a coarse-grained modeling \cite{Levitt2014} and, as an extreme case, it can be even identical with the actual microstate of the system, $\Gh =\{\xx\}$, which means that an observer has a complete information about the system.  Information about the system  can be also formed by all quantum states $|\psi_i\rangle$ in a quantum projector $\sum |\psi_i\rangle\langle\psi_i|$ describing an incomplete knowledge about a system in a mixed state \cite{StrasWint2021}. A special case of IRF  is the IRF$_{mic}$ defined by  $\Gh_{mic}(t)=\{\xx (t)\}$. 
 
\begin{figure}[h]
\includegraphics[width=0.4\textwidth]{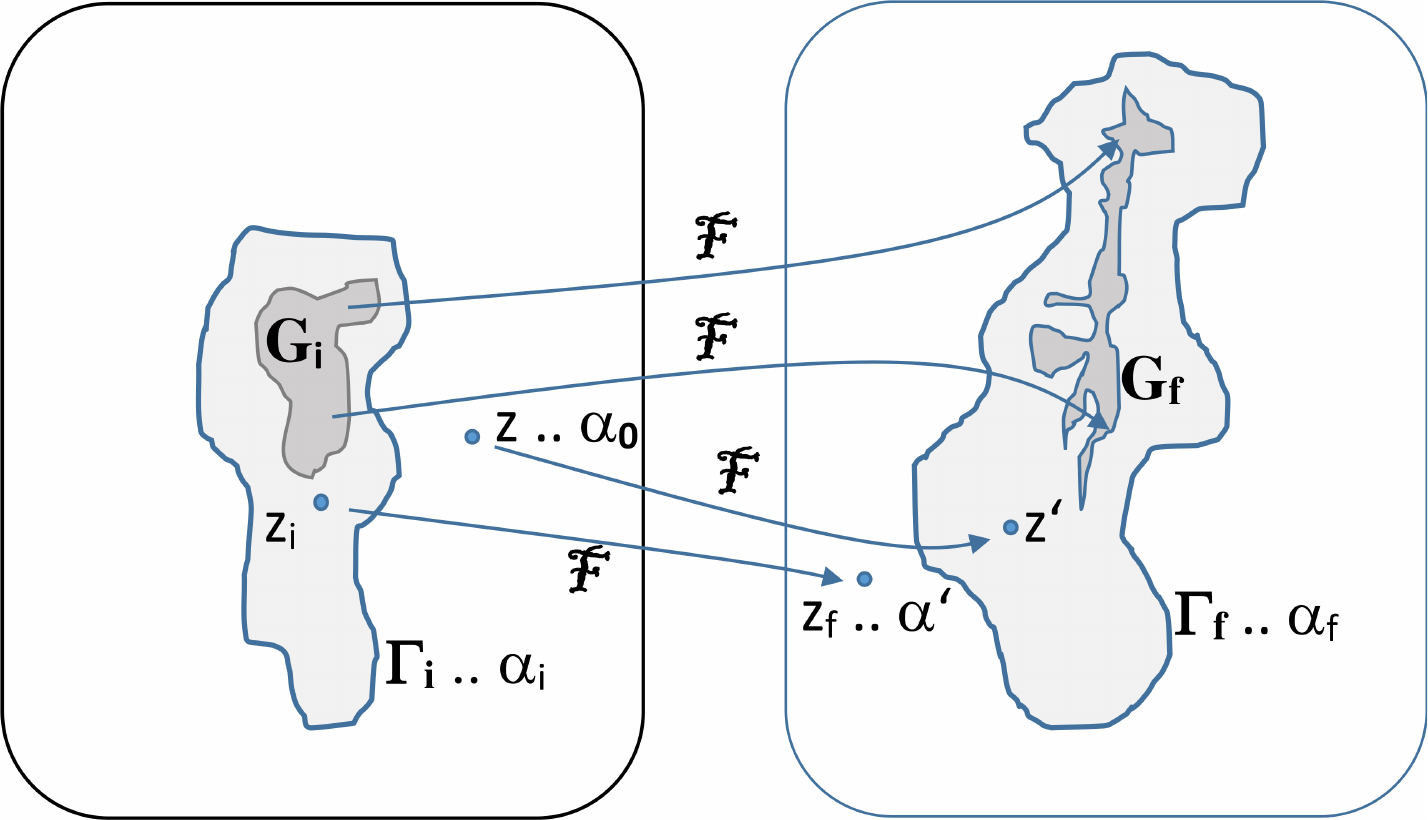}
\caption{The mapping $\fun$ represents the evolution of  states from a large state space  to their values after the time $\tau =\tF-\tI$. The change  $\ywI\to\ywF$ happens only if the initial state $\xxI\in\GsI\subset\GhI$. A state $\xaI\in\GhI$ not belonging into $\GsI$ evolves to a microscopic state $\xaF$  that corresponds to  another observable indicators, $\yw '\neq\ywF$. Similarly, a state $\xa '\in\GhF$ not belonging into $\GsF$ is a result of an observed process $\yw_0\to\ywF$ that cannot begin at $\ywI$. The sets $\Gh (t)$ define the IRF given by knowledge of current values of $\yw$ at individual times, the sets $\Gs (t)$ define the IRF'   based on the knowledge of the process $\ywI\to \ywF$. }
\label{fig:2}
\end{figure}

 The important point of the presented approach is that  referential sets (and IRF) may not be connected with any observed indicators. An important example of such a situation arises if the microscopic evolution of the system is deterministic. Let us denote $\fun$ the evolution operator $\xxF =\fun (\xxI)$, where $\xxI$, $\xxF$ is the initial and final microstate, respectively. 
Let $\GhI$, $\GhF$ be the referential sets at times $\tI$, $\tF$, respectively. Denote $\GsI\subset\GhI$ the set  of all states $\xx\in\GhI$  so that $\fun (\xx  )\in\GhF$, and  $\GsF\equiv \fun (\GsI )$ (see Fig.~\ref{fig:2}). 

We define the so-called \emph{dynamic} information reference frame IRF' as $\Gh '(\tI )=\GsI$, $\Gh '(\tF )=\GsF$. The referential sets here cannot be defied by current values of some observable indicators. If $\Gh (t)$ are defined by current values of observable indicators $\yw$ the sets $\Gs (t)$ are given by knowledge of the process $\ywI\to\ywF$. The transformation of the reference frame into the dynamic one,  IRF~$\to$~IRF', plays important role in what follows.

 \emph{Entropy}. --- From an information theoretic perspective,  entropy is associated with observer's ignorance (lack of information) as to the microstate of the system \cite{Jaynes1957,Crooks1999}. In other words,  the observer has some (incomplete) information about the microstate of the system and entropy measures the value of additional information that is necessary to "add" to this knowledge to determine the actual microstate precisely. 
 The knowledge of observer is identified here with referential information. It implies that the \emph{information entropy} is defined in an IRF as 
\begin{equation}\label{SI}
 \SI_{\Gh} (\xx )=k_B\info_{\Gh} (\{ \xx \}),
\end{equation}
where $k_B$ is Boltzmann's constant and $\Gh$ is a referential set from this IRF. 
 
 From thermodynamic point of view, entropy is a \emph{state} quantity  \cite{Callen1985,LiebYngvason2013} what means that $\SI$ can depend \emph{only} on such information that concerns the system at  a concrete time.   It implies that each referential set from an IRF must be interpreted as information about the system at a concrete time moment only.  If a set of the dynamic IRF defined above, i.e. $\Gs (t)$, is used to define entropy it  must be interpreted as special information about a concrete time moment that is not determined by  observable indicators.

The information entropy defined by Eq.~(\ref{SI}) may be understood as an objective, microscopic quantity (it depends on $\xx$) whose meaning is simply 'information about the actual microstate'. The well-known doubts concerning objectivity  of  entropy become irrelevant here because    
its observer-dependence is nothing else but a usual dependence of any physical quantity on some reference element (unit, frame, etc.). The concrete value of entropy is the value of information evaluated always in a concrete IRF. Notice that the entropy is, for example, always zero in the IRF$_{mic}$.

\emph{Properties of information value}. ---
Consider
 a physical phenomenon that can happen if and \emph{only} if $\xx\in\om$. A very large value of $\info_\Gh (\om )$ means that the  realization of this phenomenon is very rare for this observer (the value of message "this number will win in lotto" is very high).  Hence $\info_\Gh (\om )\to\infty$   means that the observer can observe this phenomenon with a overwhelmingly low  probability, i.e. practically cannot observe this phenomenon at all. 
 
Let us postulate the two important properties of $\info_\Gh$: 
\\
(i) If $\om\subset \Gh_2\subset \Gh_1$ then the observer who knows $\Gh_1$ needs additional information $\xx\in \Gh_2$ to become equivalent to that one who knows $\Gh_2$, i.e. we postulate
\begin{equation}\label{I1}
\info_{\Gh_1}(\om ) = \info_{\Gh_2 } (\om )+\info_{\Gh_1} (\Gh_2).
\end{equation}
(ii) The smaller is the set $\om$ with respect to $\Gh$ the higher must be the value of information detecting $\om$. That is why we postulate
\begin{equation}\label{inf0}
\info_{\Gh} (\om )=\F_\Gh (M_{\Gh} (\om )) \ge 0,
\end{equation}
where   $\F_\Gh$ is  a decreasing real function and $M_\Gh$ is a real measure of  $\om$ on $\Gh$ so that $\info_\Gh (\Gh)=0$.  
Concerning the measure $M_\Gh$ we suppose  that $M_\Gh (\om )\to 0$ if the size of $\om$ goes to zero. It implies \cite{Rudin1987} that $M_\Gh$ is expressed via a real, non-negative function $f_\Gh$ so that
$M_\Gh (\om )=\int_{\om}f_\Gh (\xa ){\rm d}\xa$ in continuous, and $M_\Gh (\om )=\sum_{\xa_i\in\om}f_\Gh (\xa_i )$ in discrete cases.

The special case of $\info_\Gh$ fulfilling the postulated properties is  $\info_\Gh^{Sh}(\om ) =-\ln p(\om| \Gh )$ (Shannon's formula), where $p(\om| \Gh )$ is the probability measure on $\Gh$, i.e. $M_\Gh (\om )=p(\om| \Gh )$, $\F_{\Gh}=-\ln$. The information value $\info_{\Gh}$, however, can be defined without any connotation to probabilities.

\section{Adiabatic processes}
\label{sec:AD}

Any process governed only by inner dynamics of an isolated system is \emph{adiabatic}. The concept of adiabatic processes is, however, broader. Namely,  it is any process during which  the exchange of energy with its surrounding exists only in the form of work. Work is defined in classical macroscopic thermodynamics, its generalization to meso- or microscopic processes is far from being easy and straightforward. 

The well-accepted, scale-independent  definition of processes during which the system-surrounding interaction is realized only with the exchange of work concerns situations when  the evolution of observable indicators is firmly given by a prescribed protocol \cite{Jar1997,Crooks1998,Crooks2000}
and no other external intervention exists. Though there exists an exchange of energy with the system environment,  such a process is fully deterministic and reversible. It motivates us to \emph{define} the adiabatic process as \emph{any} deterministic and reversible process $\xxI\to\xxF$, where $\xx$ is the full (microscopic) state of the system.  

\begin{figure}
\includegraphics[width=0.4\textwidth]{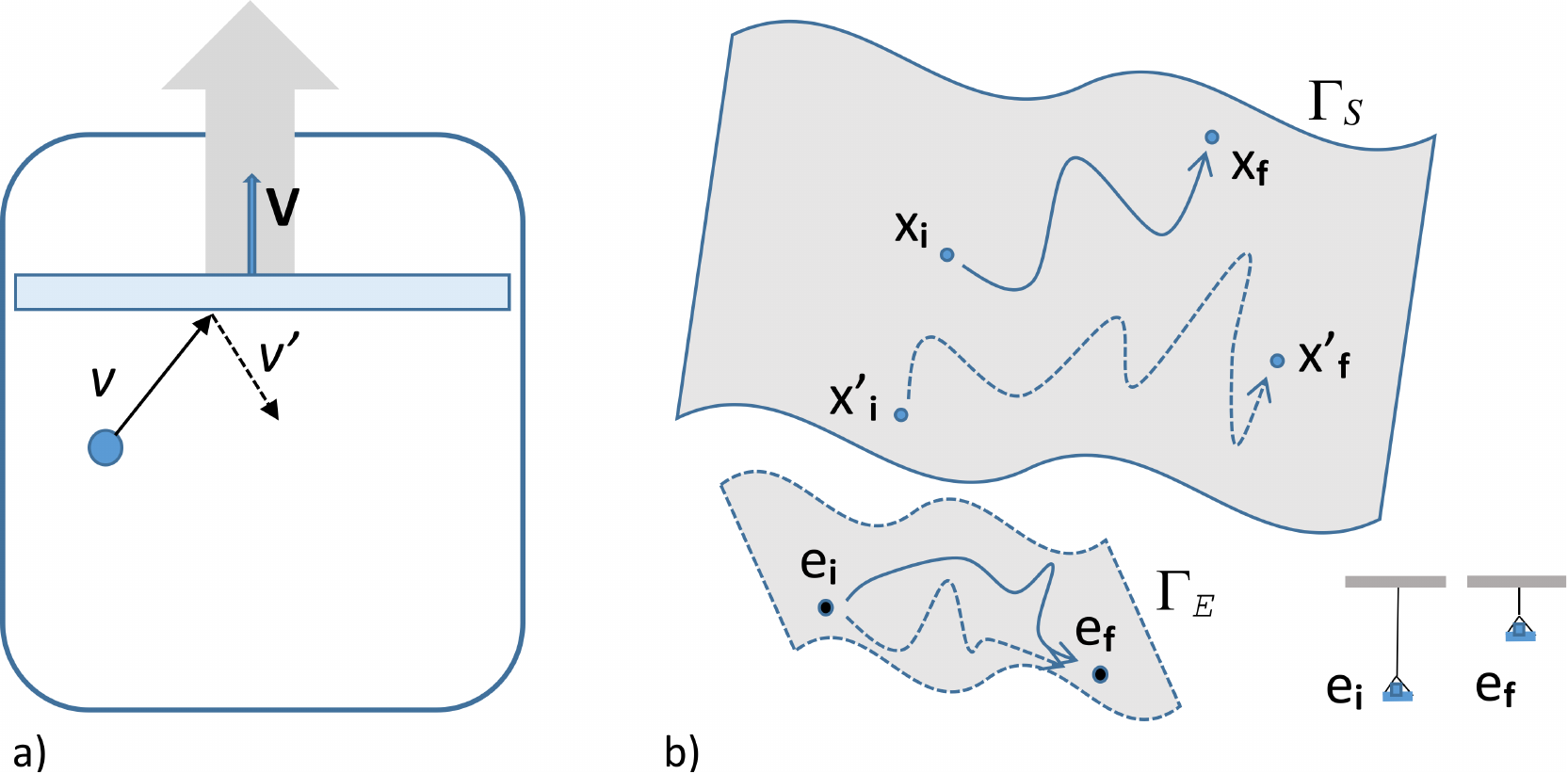}
\caption{Two examples of an adiabatic process. a) The piston moves with an externally controlled velocity $V$. A molecule elastically impacting this piston changes own kinetic energy, $\Delta E_{kin}= -2mV(v_z-V)$, where $v_z$ is its perpendicular velocity, $v_z> V$, and $m$ its mass. This change of energy is interpreted as the \emph{supplied work}. No transfer of energy into the piston can be described here. b) The system $\syst$ interacts with its environment $\env$ so that during any process the environment is at given states $e_i$ and $e_f$ at the beginning and the end of the process, respectively. These states define the energy transformed from $\syst$ to $\env$, $\Wen (e_f,e_i)$, e.g. the change of a weigh in gravity field. The processes are deterministic and reversible.       }
\label{fig:3}
\end{figure}

A concrete realization of the 'prescribed-protocol' interaction means the existence of macroscopic massive bodies or some device that are insensitive to microscopic interactions. The work then can be defined  for the system alone (supplied work) while its transfer into the surrounding cannot be quantify anyhow (see Fig.~\ref{fig:3}a) \cite{MaesTasaki2007,Peliti2008,Holecek2019}.   The above definition of the adiabatic process as any deterministic and reversible (unitary) process, however, does not exclude possibility of defining the work performed in the surrounding.   

   Let $\syst$ be a system and we study processes beginning at various microscopic states of this system, $\xxI\in\GA$. During any process, the system interacts with its environment, $\env$, so that $\syst +\env$ is a fully isolated system (universe). Suppose that the environment and the system-environment interaction is designed in such a way that the initial (microscopic) state of the environment is always $e_i$ and its final state (say, at a sufficiently large time) is always $e_f$. The transferred work into this environment, $\Wen$, is a function of $(e_f,e_i)$ (see Fig.~\ref{fig:3}b).

      The universe $\syst +\env$ is an isolated system, hence the process $(\xxI ,e_i)\to (\xxF ,e_f)$ is deterministic and reversible. Since $e_i$ and $e_f$ are fixed, the process $\xaI\to\xaF$ is deterministic and reversible too (though the trajectory of the environment and its interaction with the system is different for various initial conditions of the system). This idea is a generalization of  the concept of adiabatic accessibility  \cite{LiebYngvason1999} into microscopic scales. 

\emph{Information conservation}. ---
During adiabatic  evolution, information about  possible microstates of the system has to be conserved. The reason is that the evolution is deterministic and reversible which means that the configuration at any time moment carries information about that at any other time. 

Let $\fun_\tau$, $\tau\in\bf{R}$ be a set of one-to-one mappings expressing the time shift of the system microstate  during  time evolution, i.e. $\xx (t+\tau )=\fun_\tau (\xx (t))$ (hence the inverse mapping of $\fun_\tau$ is $\fun_{-\tau}$). Since the value of information about microstates is always related to  referential information the  \emph{law of conservation of information}  must be  expressed as follows:  
\begin{equation}\label{CI}
\forall \tau\, \forall\om\subset\Gh :\,\,\, \info_{\fun_\tau (\Gh)}(\fun_\tau (\om )) =\info_{\Gh }(\om ).
\end{equation}
The referential set, however,  usually does \emph{not} vary according the rule $\Gh\to\fun_\tau (\Gh )$. If it is defined by values of some observable indicators, i.e. $\Gh(\yw )$, the set $\Gh (\yw (t+\tau ))$ is usually different from $\fun_\tau (\Gh (\yw ))$  (see Figs.~\ref{fig:1},\ref{fig:2}).

\emph{Adiabatic entropy change}. --- The transformation into a dynamic reference frame IRF', IRF~$\to$~IRF', can be written by the use of  Eq.~(\ref{I1}). If $\om\subset\GsI$ we get
$$\info_{\GhI} (\om )=\info_{\GsI} (\om )+\info_{\GhI} (\GsI ),$$
and 
$$\info_{\GhF}(\fun_\tau (\om ) )=\info_{\GsF }(\fun_\tau(\om ))+\info_{\GhF} (\GsF ).$$
Since $\GsF =\fun_\tau (\GsI )$ the law of conservation of information cancels the first terms on the right-hand sides of the equations and we get that 
$\info_{\GhI}(\om ) -\info_{\GhF}(\fun_\tau (\om ) )$ does not depend on $\om$, i.e.
\begin{equation}\label{It} 
\info_{\GhF}(\fun_\tau (\om ) )-\info_{\GhI}(\om )=\info_{\GhF}(\GsF )-\info_{\GhI}(\GsI )  \equiv \Delta\infJ,
\end{equation}
where $\Delta\infJ = \infJ '-\infJ$, $\infJ \equiv \info_{\GhI}(\GsI  )$, $\infJ '\equiv \info_{\GhF}(\GsF)$. 

If we chose $\om =\{\xx\}$, $\xx\in\GsI$, we get the  change of  the entropy distribution  $\Delta\SI (\xx )=\SI_{\GhF}(\fun_\tau (\xx ))- \SI_{\GhI }(\xx )$ so that 
\begin{equation}\label{entropy}
 \Delta\SI (\xx )=  k_B\Delta\infJ .
\end{equation} 

The right-hand side of Eq.~(\ref{entropy}) depends only on $\GhI$, $\GhF$ and the mapping $\fun_\tau$. It implies that the change of entropy is the same for each $\xx\in\GsI$,
\begin{equation}\label{scaleS}
\forall \xx ,\xx' \in\GsI:\,\, \Delta\SI (\xx )=\Delta\SI (\xx ' )\equiv \Delta\SI .
\end{equation}
The change of entropy during an adiabatic process that  
is observed as $\ywI\to\ywF$ must begin at $\GsI$ hence $\Delta\SI$ is determined by the change of $\yw$, i.e. $\Delta\SI (\ywI\to\ywF )$.
It is a very important result: though the information entropy is a microscopic quantity its change is determined only by the change of quantities defined at the scale where the system is studied. 

\section{Concrete form  of the information entropy}
\label{sec:GB}

Let $\Gh$ be a subset of a system  state space and $\xx\in\Gh$. The value of information entropy $\SI_\Gh (\xx )$ is uniquely determined by  $\xx$ and $\Gh$. 
To find out this dependence we suppose that there exists a sufficiently small $\tau_0$ so that for each $0< \tau < \tau_0$ the element $\fun_\tau (\xx )\in\Gh$, where $\fun_\tau$ represent an adiabatic process. Hence we can choose the  same referential sets, $\Gh (t)=\Gh$, for each $t\in\langle 0,\tau_0\rangle$. It allows us to form the IRF for each $\tau$ ($\tau<\tau_0$) so that    $\Gh (0 ) \equiv \GhI =\Gh$ and $\Gh (\tau )\equiv\GhF =\Gh$. 

Since $\Gh$ is fixed in this consideration we omit the index $\Gh$ in all quantities, i.e. $\SI \equiv\SI_\Gh$ and the quantities in Eq.~(\ref{inf0}):  $M\equiv M_{\Gh}$, $f\equiv f_\Gh$, and $\F\equiv\F_{\Gh}$. 
   Since $\SI (\xx )=k_B\info_\Gh (\{\xx \})=k_B\F (f(\xx ))$ we get 
\begin{equation}\label{proof1}
f(\xx )=\F_{-1}(\QQ (\xx )),
\end{equation}   
where $\F_{-1}$ is the inverse function of $\F$ and  $\QQ (\xx )\equiv k_B^{-1}\SI (\xx )$.

Suppose that the state space is discrete. Then
the measure $M(\Om )=\sum_{\xx\in\Om}f(\xx )$ and  Eq.~(\ref{It}) becomes
\begin{equation}\label{proof2}
\F (\sum_{\xx\in\Om}\F_{-1}(\QQ (\xx ) +\Delta\infJ ))=\F (\sum_{\xx\in\Om}\F_{-1}(\QQ (\xx )))+\Delta\infJ 
\end{equation}
(we use Eq.~(\ref{entropy}) to express that $\QQ (\fun (\xx ))=\QQ (\xx )+\Delta\infJ$).

 The validity of Eq.~(\ref{proof2}) for all $\tau\in (0,\tau_0)$ determines   a possible form of  $\info_\Gh$ (see Appendix), namely
\begin{equation}\label{IP1}
\info_{\Gh} (\om )= -\del \ln \sum_{\xx\in\om }\e^{-\del^{-1} k_B^{-1} \SI_\Gh (\xx )}, 
\end{equation}
where $\del$ is a positive constant. It is easy to check that this form of $\info_\Gh$ fulfills Eq.~(\ref{It}) for any $\GhI$, $\GhF$ and $\Delta\infJ$ if  the entropy distribution fulfills  Eq.~(\ref{entropy}).  

To fulfill the condition $\info_\Gh (\Gh)=0$  
the entropy has to have a form 
\begin{equation}\label{entS}
\SI_\Gh (\xx )= k_B\del (\ln |\Gh|_\Q +\Q_\Gh (\xx ) ),
\end{equation}
where $\Q_\Gh$ is a real function on the state space of the system that may depend also on $\Gh$, $\del$ is a constant, and $|\Gh|_\Q$ is the special measure of the set $\Gh$,
\begin{equation}\label{Ssum}
|\Gh|_\Q= \sum_{\xx\in \Gh }\e^{- \Q_\Gh (\xx )}.
\end{equation}
 The sums $\sum_{\xx\in\Om}$ in Eqs.~(\ref{entS},\ref{Ssum}) must be  replaced by integrals, $\int_\om{\rm d}\xx$, in the case of the continuous state space. The constant $\del$ only rescales the information value and information entropy. In what follows we fix  $\del =1$. 
 
The distribution of $\Q_\Gh$ on $\Gh$ depends on the mappings $\fun_\tau$. Namely putting Eq.~(\ref{entS}) into Eq.~(\ref{It}) we get
\begin{equation}
\Delta\Q (\xx )\equiv \Q_{\GhF}(\xx (t+\tau ))-\Q_{\GhI} (\xx (t))= \Delta\infJ (\tau) +\ln\frac{|{\GhF}|_\Q}{|{\GhI}|_\Q}.
\end{equation}  
It implies that  
\begin{equation}\label{DPhi}
\forall\xx\in\GsI (\fun ) :\,\,\, \Delta\Q (\xx )=C(\GhI ,\GhF,\fun ).
\end{equation}
This condition is analogical to Eq.~(\ref{scaleS}) which means that  the change of function $\Q$ is similarly defined on the observation scale.

The only entropy distribution that fulfills Eq.~(\ref{entropy}) for an arbitrary $\fun$ arises if we choose $\Q_\Gh (\xx )= C(\Gh )$, i.e. it does not depend on $\xx$. We get  
\begin{equation}\label{Bent}
\SI_\Gh (\xx )= k_B\ln |\Gh | =S_B,\,\,\,\, (\xx\in\Gh )
\end{equation}
where $|\Gh|$ is the volume of the set $\Gh$. That is we get the Boltzmann entropy $S_B$ as a special case of the information entropy (\ref{entS}).

\emph{Probabilistic interpretation}. --- 
The formula (\ref{IP1}) can be interpreted otherwise.  If we define  
\begin{equation}\label{defpr2}
\pr_\Gh (\om ) \equiv\e^{-\del^{-1}\info_\Gh(\om )}>0.
\end{equation}
and 
write simply $\pr_\Gh (\xx )$ instead of $\pr_\Gh (\{\xx\})$, Eq.~(\ref{IP1}) can be written as
\begin{equation}\label{defpr1}
\pr_\Gh (\om )=\sum_{\xx\in\Gh}\pr_\Gh (\xx ),
\end{equation}
and $\sum_{\xx\in\Gh} \pr_\Gh (\xx )=1$ because of the condition $\info_\Gh (\Gh )=0$. 

The values of $\pr_\Gh$ thus can be understood as a probability distribution over the referential set $\Gh$ so that the probability 
\begin{equation}\label{PPP}
P(\xx\in\om )=\e^{-\info_\Gh (\om )}
\end{equation}
when putting $\del =1$.
It implies the Shannon-like expression of entropy,
\begin{equation}\label{prS}
\SI_\Gh (\xx )= - k_B\ln \pr_\Gh (\xx ).
\end{equation}  
Notice that Eq.~(\ref{PPP}) implies that $\info_{\Gh} (\om )\to \infty$ if and only if $P(\xx\in\om)\to 0$.

\section{The second law of thermodynamics}
\label{sec:SLT}

Consider an observer, say Alice, who knows data about a system $\syst$ (possibly meso- or microscopic) only at a given time $\tI$: she  performs a measurement at $\tI$ and gains some data $\yw (\tI )=\ywI$ about the system. Alice knows that the next evolution of $\syst$ is adiabatic (its microscopic evolution is deterministic and reversible). Alice asks which value of $\yw$ can be detected at a time $\tF >\tI$. 

Each possible value of $\yw (\tF )\equiv \yw_\star$ forms a subset $\Gs (\yw_\star )\subset\GhI$ of all possible initial microstates  that realize the evolution $\ywI\to\yw_\star$ (this subset may be empty). Alice does not know, however, in which subset the actual microstate occurs. Nevertheless, information that $\xxI\in\Gs (\yw_\star )$  has different values for various $\yw_\star$, namely $I(\yw_\star )=\info_{\GhI}(\Gs (\yw_\star ) )$. The largest this value the smaller probability $P$ that $\xxI$ belongs into this subset as implied by Eq.~(\ref{PPP}),
\begin{equation}\label{P}
P(\xxI\in\Gs (\yw_\star ))= \e^{-I(\yw_\star )}.
\end{equation}
Hence if $I(\yw_\star )\to\infty$ then the result $\yw (\tI )=\yw_\star$ cannot occur since its probability is zero.

In macroscopic thermodynamics, the second law can be expressed as impossibility of adiabatic processes during which the entropy decreases, i.e. during which $\Delta\SI < 0$. Let us study the situation when the entropy decreases by the use of Eq.~(\ref{entropy}) in the macroscopic limit $k_B\to 0$ so that $\Delta S\neq 0$. 

 If $\Delta \SI/{k_B}\to -\infty$, Eq.~(\ref{entropy}) implies  $\infJ -\infJ '\to \infty$, i.e. $\infJ \to\infty$ since $\infJ '\ge 0$. $\infJ=\info_{\GhI}(\GsI )=I(\ywF )$ is  the value of information that $\xxI\in\GsI$, i.e. that  the process $\ywI\to\ywF$ will happen.   
 Eq.~(\ref{P}) implies that if $\Delta \SI <0$ then the probability of this process is zero. Hence only macroscopic adiabatic processes with non-decreasing entropy are possible. Eq.~(\ref{entropy}) thus expresses the second law of thermodynamics. 
\begin{figure}
\includegraphics[width=0.4\textwidth]{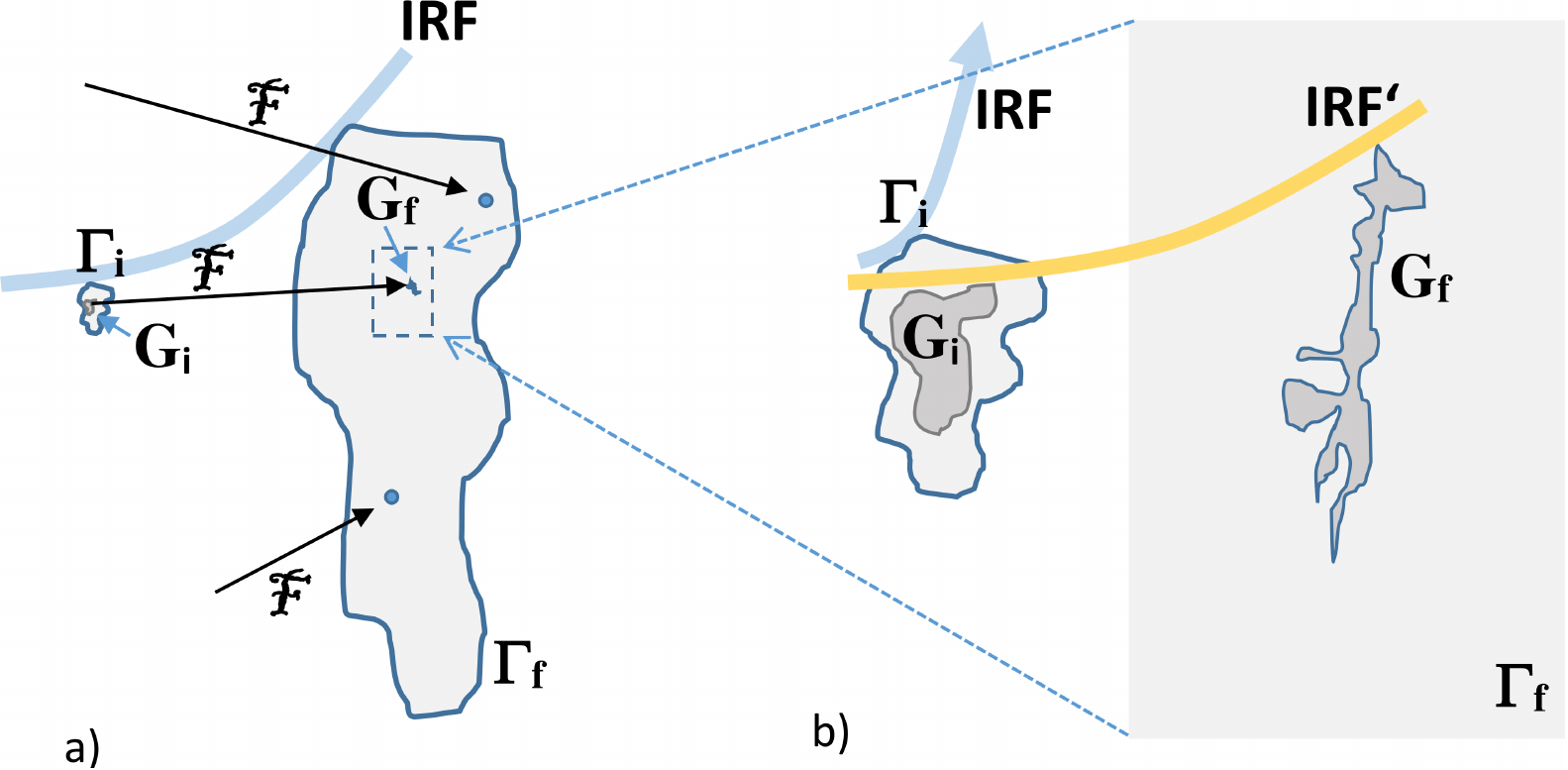}
\caption{Consider a process $\ywI\to\ywF$ from $\tI$ to $\tF$ so that  the set $\GhF$ corresponding to $\ywF$ is overwhelmingly larger than the set $\GhI$ corresponding to $\ywI$. It is a typical macroscopic situation when $\Delta S/k_B$ is extremely  large, say $\sim 10^{20}$. a) If an observer, Bob,  has information \emph{only} concerning $\tF$, i.e. he knows $\ywF$, he describes the process at the IRF given by sets $\Gh (t)$ and he  must infer that the probability that the origin of the situation at $\tI$ is a point from an extremely small region $\GhI$ is negligibly small, $P\sim 10^{-10^{20}}$. He must conclude that $\yw (\tI )$ cannot be $\ywI$.
b) If Bob has information (a record) that $\yw$ at $\tI$ was $\ywI$ then he has information that corresponds to the description in IRF' defined by sets $\Gs (t)$.  }
\label{fig:4}
\end{figure}

 This formulation of the second law is formally very close to Boltzmann's statistical derivation of this law. 
The main argument is probabilistic too: if $\Delta S<0$ and $|\Delta S| \gg k_B$ the "target" set $\GsF$ is extremely smaller than the set of all initial possibilities ($\GhI$) and there is an overwhelmingly small probability of "hitting" it (to realize the process $\ywI\to\ywF$) \cite{Lebowitz1993,Penrose2005}. 

The information formulation, however, can avoid the principle  problem of Boltzmann, i.e. the fact that the statistical argumentation can be used in the opposite time direction to make a paradox \cite{Penrose2005,Albert2000,Earman2006,Callender2021}.
To show it, imagine another observer, say Bob, who has information about the same system $\syst$ at time $\tF$ without communicating with Alice. Bob detects the value $\yw (\tF )=\ywF$ and asks which value of $\yw$ was at time $\tI <\tF$. If $\Delta \SI =\SI (\tF )-\SI (\tI )> 0$ we get from Eq.~(\ref{entropy}) in the macroscopic limit that $\infJ '\to\infty$. The probability that $\xxI$ was in $\GhI$ is thus overwhelmingly small (see Fig.~\ref{fig:4}a). Bob must conclude that the past of the system could not be so that $\yw (\tI)$ was $\ywI$. This conclusion is, however, false.

Nevertheless, the situation of Alice and Bob  is not symmetric from information point of view. Bob can receive a message from Alice about the situation at $\tI$ while Alice cannot have a message from Bob about the situation at $\tF$ (information cannot be send into the past). The conclusion of Alice is thus based on \emph{all} possibly attainable information (at her observation scale) about the system at $\tI$. The situation of Bob is different. There are two possibility concerning his information state (see Fig.~\ref{fig:4}):\\[1mm]
(i) Bob has a \emph{record} about the situation at $\tI$ (e.g. Alice's message). Then his information about the system is not only $\yw (\tF )$ (his measurement) but also $\yw (\tI )$ (the record). Having this information Bob cannot use $\GhF$ as the set of \emph{all} possibilities corresponding to his information about the system. If he knows that $\yw (\tI )=\ywI$ and $\yw (\tF )=\ywF$ then all possibilities are given by sets $\GsI$ and $\GsF$. Bob thus uses the dynamic referential frame, IRF', in which $\Gh '(\tI )=\GsI$, $\Gh '(\tF )=\GsF$ (see Fig.~\ref{fig:4}b). In this referential frame $\Delta \SI'=0$ and Eqs.~(\ref{entropy},\ref{PPP}) in IRF' gives  that $P(\xxI\in\GsI)=P(\xxF\in\GsF )$. No paradox arises.\\[1mm]
(ii) There is no record about the situation at $\tI$. Bob concludes from Eq.~(\ref{entropy})  that $\yw$ at $\tI$ could not be $\ywI$. His conclusion, however, cannot be verified anyhow:  Bob cannot send a message into the past and no record about  the situation at $\tI$ exists. Whenever a record about the situation at $\tI$ appears (e.g. in a form of an indirect physical proof) Bob must shift his consideration into the dynamic referential frame  defined by sets $\Gh '=\Gs$ and we get  again the situation (i).

\section{$\Q$-representations of entropy}
\label{sec:SC}

 The function $\Q_\Gh$  has  a unique meaning at the observable scale since its change is determined only by the process $\ywI\to\ywF$ as implied by Eq.~(\ref{DPhi}). 
 A concrete choice of the function $\Q_\Gh$  defines  the information entropy and, consequently, the information value $\info_\Gh$. We call it the $\Q$-\emph{representation}.  
 The simplest $\Q$-representations are related to several kinds of entropy:\\[1mm]
\emph{Boltzmann entropy}. It arises in the $C$-representation in which the function $\Q_\Gh$ is constant (Eq.~(\ref{Bent})).\\  
\emph{Shannon entropy}. The probabilistic interpretation of information entropy as given by Eqs.~(\ref{PPP},\ref{prS}) can be interpreted as a special $\Q$-representation, called the $p$-representation.  We get it if the first term in Eq.~(\ref{entS}) is identically  zero, i.e. $|\Gh|_\Q =1$ for an arbitrary $\Gh$.   
If denote $\pi_{\Gh} (\xx )\equiv \exp (-\Q_{\Gh} (\xx ) )>0$, we get that the information entropy in $p$-representation, $\SI_p$, has now the form of Shannon entropy, Eq.~(\ref{prS}), $\sum_{\xx\in \Gh} \pi_{\Gh} (\xx )=1$ and $\pi_\Gh =p_\Gh$ are probabilities.  
\\ 
\emph{Clausius (equilibrium) entropy}.  
Another possible $\Q$-representation we get if 
  $\Q_{\Gh}$ is  a constant of motion, i.e. $\Delta\Q =0$. 
  If the system  energy is conserved during the studied process we can identify $\Q_{\Gh} (\xx )=\beta \ha (\xx )$, where $\ha (\xx )$ is the system hamiltonian and
 $\beta\equiv (k_BT)^{-1}$ with $T$ being an artificially chosen constant.
   In this $\ha$-representation we have
\begin{equation}\label{Hinf}
\info_{\Gh} (\om ) =-\ln \frac{\sum_{\xx\in\om }\e^{- \beta\ha (\xx )}}{\sum_{\xx\in {\Gh} }\e^{- \beta\ha (\xx )}},\,\,\,\,\,\, (\ha-{\rm rep.})
\end{equation} 
and  Eq.~(\ref{entS}) becomes 
\begin{equation}\label{Hent}
\SI_{\Gh} (\xx ) =T^{-1}(\ha (\xx )- F_{\Gh})=S_C,\,\,\,\,\,\, (\ha-{\rm rep.}) 
\end{equation} 
 where $F_{\Gh}\equiv  -k_BT\ln \sum_{\Gh} \exp (-\beta \ha (\xx ))$ is the free energy. If $\xx\in\Gh_{eq}$, where $\Gh_{eq}$ is  the equilibrium macrostate of the system, the hamiltonian can be identified with the system internal energy, $U =\ha (\xx )$, the parameter $T=T_{eq}$ is the thermodynamic temperature,  and the information entropy becomes the classical (Clausius) entropy $S_C$. 

\section{Probabilities}
\label{sec:prob}

 When assuming that the entropy in $\Q$-representation equals the entropy in $p$-representation, i.e. $\SI_{\Gh} =S_p$, we find the probabilities in individual representations. In $C$-representation we get 
\begin{equation}\label{pBolt}
p_{\Gh}(\xx) =|\Gh|^{-1}.\,\,\,\,\,\, (C-{\rm rep.}) 
\end{equation} 
Similarly in $\ha$-representation we have 
\begin{equation}\label{peq}
p_{\Gh}(\xx )=\frac{\e^{-\beta\ha (\xx )}}{Z_{\Gh}} =p_{eq}(\ha(\xx )),\,\,\,\,\,\, (\ha-{\rm rep.})
\end{equation}
where $Z_{\Gh}=\sum_{\Gh}\e^{-\beta\ha (\xa )}$. It means that the probability  of a microstate corresponds in the $\ha$-representation with the Boltzmann-Gibbs equilibrium distribution, $p_{eq}(\ha )$.  
 Its  interpretation  is, however, different. The microscopic energy of a system in thermal contact with a reservoir fluctuates and reaches the energy ${\cal E}$ with the probability $p_{eq}({\cal E} )$. We, however, study an isolated system, i.e. $\ha (\xx (t))$ is constant along its trajectory. The probability $p_{\Gh}(\xx )$ expresses in $\ha$-representation nothing but observer's knowledge as to $\xx$ in dependence on $\Gh (t)$ (that is expressed through $Z_{\Gh}$).  

During the unitary evolution the entropy change in $p$-representation, $\Delta\SI =  -(\ln p_{\GhF}(\fun(\xx ))-\ln p_{\GhI}(\xx ))$, must be the same for all $\xx\in\GsI$, i.e.
\begin{equation}\label{pp}
\pout =\pin\e^{-k_B^{-1}\Delta\SI},
\end{equation} 
where $$\pin =\sum_{\xx\in\GsI} p_{\GhI}(\xx ),\,\,\, \pout= \sum_{\xx '\in\GsF} p_{\GhF}(\xx ').$$
This result looks strange. Namely we expect that the probability of occurrence of the system at a (micro)state $\xx$ so that  the process $\ywI\to\ywF$ is realized, i.e. $\pin$, must be the same as the probability of finding the final (micro)state that corresponds to the fact that the process $\ywI\to\ywF$ has been realized, i.e. $\pout$. However,  if $\Delta\SI \neq 0$ then $\pin\neq\pout$. 

The explanation of this (and other) seeming inconsistencies connected with Eq.~(\ref{pp}) consists in understanding the entropy in  information context. Namely information entropy is given only by information about the system concerning only one time moment. A correct interpretation of Eq.~(\ref{pp}) should be done via the idea of two independent observers so that the first has information about the system at $\tI$ only, the second does the same at $\tF$ (Alice and Bob in Section~\ref{sec:SLT}). The first one observes $\ywI$ and can (in principle) deduce that the process $\ywI\to\ywF$ happens with the probability $\pin$. The second one observes $\ywF$ and can (in principle) deduce that the process $\ywI\to\ywF$ has happened with the probability $\pout$ (she/he does not know the past of the system).    

If an observer knows that the process $\ywI\to\ywF$ has happened this observer  knows more than the previous ones, i.e. she/he  knows that $\xxI\in\GsI$ and $\xxF\in\GsF$. We can  describe  this situation in the dynamic information reference frame,  IRF'$=\{\Gh '(t)\}$, in which the referential sets $\Gh '(\tI )=\GsI$ and $\Gh '
(\tF )=\GsF$. In the IRF', $\Delta\SI ' =0$  and the corresponding probabilities $\pin '$ and $\pout '$ are now identical. 

A direct connection of information entropy and probabilities shows the necessity of defining also probabilities with respect to an information reference frame, IRF. The interpretation of 'probability of an event' thus depends on a chosen  IRF too. It is the context in which Eq.~(\ref{pp}) must be interpreted.

\section{Information form of the generalized second law}  
 \label{sec:TD}

The expression of entropy in $\ha$-representation,  Eq.~(\ref{Hent}), can be interpreted in standard thermodynamics concepts in the case of a macroscopic  system  in thermal equilibrium. In this Section, we use the $\ha$-representation to find the thermodynamic interpretation of the information entropy for an arbitrary (micro-, meso-, macroscopic) system in a general nonequilibrium state.
 
We study a more complex structure called here the supersystem that is perfectly isolated from surroundings so that its evolution is deterministic and reversible and its energy is constant. The observable indicators of the supersystem are its special degrees of freedom, $\yw$, so that the complete state of the supersystem   $\xx =(\st ,\yw )$. The studied thermodynamic system is then defined by the (internal) degrees of freedom $\st$.  
    We can  write 
\begin{equation}\label{divH}
\ha (\xx )= \E (\yw )+\ha_{int} (\st ,\yw ),
\end{equation} 
  where     $\E (\yw )$ is the  energy connected with the observable degrees of freedom 
 and $\ha_{int}$ represents the rest of energy connected with internal degrees of freedom and their interactions with the observed ones. The energy of the studied thermodynamic system is identified as $\ha_{int}$. 

A simple example is presented at Fig.~\ref{fig:1}. The supersystem is the whole structure including the gas and the piston. The position and velocity of the piston represent the observable indicators so that    
   $\E (\yw ) =MgX + 1/2 MV^2$ with $M$ being the mass of the piston.  The gas is the own thermodynamic system. The change of $\E$ during a process measures the change of energy of the gas.

We study processes during which $\Delta\ha = 0$ (the supersystem is isolated).  
  The change of information entropy in $\ha$-representation during the single (microscopic) process $\xxI\to\xxF$ is given by Eq.~(\ref{Hent}),   
\begin{equation}\label{1TD}
T\Delta\SI =\Delta\ha_{int}  -\Delta F(\yw ,T),\,\,\,\,\,\, (\ha-{\rm rep.})
\end{equation} 
where $F$ corresponds to the free energy of the internal degrees of freedom,
\begin{equation}\label{F}
F(\yw ,T)=-k_BT\ln \sum_{\st'\in \Gh(\yw )}\exp (-\beta \ha_{int}(\st ',\yw )).
\end{equation}
 During the  process  the energy $\W =\Delta\ha_{int} =-\DE$ (since $\Delta\ha =0$) is  transferred from observable degrees of freedom into the internal ones, i.e. into the thermodynamic system. This energy transfer can be called the \emph{supplied work} regardless if the system is macroscopic or microscopic  \cite{Jar2004}.  
 The quantity
  $T\Delta\SI$ thus can be identified with the \emph{dissipated work},  
 $\W_{dis}=\W -\Delta F$, defined for a single trajectory 
 \cite{KavParBro2007}, 
\begin{equation}\label{dis}
 T\Delta\SI =\W_{dis}.\,\,\,\,\,\, (\ha-{\rm rep.}) 
\end{equation}
The parameter $T$ is an arbitrarily chosen positive real number. Its value can be fixed and identified with a thermodynamic temperature with using the idea of contact temperature \cite{Muschik2021}. 

Namely if an arbitrary system (micro-, meso-, macroscopic) is at the state $\xx_0 = (\st_0 ,\yw_0 )$ we can put it into a thermal contact with a sufficiently large thermal reservoir in equilibrium with the  temperature $T_{eq}$,  fix the value of $\yw$ and keep relaxing the internal degrees of freedom into thermal equilibrium, i.e. $(\st_0 ,\yw_0 )\to (\st_{eq},\yw_0 )$. During the relaxation the energy $\Qh_{rel} =\Qh (\st_0 ,\yw_0 ,T_{eq})$ is absorbed by the internal degrees of freedom. In dependence on $T_{eq}$ this energy may be  positive or negative. We can choose $T_{eq}$ so that $\Qh_{rel} =0$ and identify the parameter $T$ with this temperature, $T=T_{eq}$. The temperature  $T$ thus depends on a chosen state of the system, i.e. $T=T(\st_0,\yw_0 )$.

 \emph{Generalized second law}. ---
The change of energy $\E$ connected with observable degrees of freedom $\yw$ is the maximal energy that can be used by the observer who detects (and principally can control) these variables, i.e. the \emph{usable work}, $\W_u =-\W =\Delta\E$. If $\Delta\SI \ge 0$ we get from Eq.~(\ref{1TD}) the familiar inequality of classical thermodynamics,
\begin{equation}\label{ineTD}
\W_{u}\le -\Delta F.
\end{equation}
We can express the relation between $\W_{u}$ and $\Delta F$ more precisely in a full general situation by using the transformation of the used IRF (that is defined by observable indicators $\yw$) into the dynamic IRF' with $\Gh '(t)=\Gs (t)$. In the IRF'  we can define  the free energy, $F'$, and the entropy $\SI '$, and we get  $\Delta\SI ' =0$ and Eq.~(\ref{1TD}) gives $\Delta\E =- \Delta F '$. Using Eq.~(\ref{Hinf}) we get $\Delta F' =\Delta F +\Delta\infJ$, i.e.
\begin{equation}\label{WF}
W_{u} = -\Delta F - k_BT\Delta\infJ .
\end{equation}    
Since $\Delta\infJ =\infJ '-\infJ$ and $\infJ '\ge 0$ we get the general inequality 
\begin{equation}\label{WF2}
W_{u} \le -\Delta F + k_BT\infJ ,
\end{equation}    
that generalizes the standard inequality Eq.~(\ref{ineTD}) since $\infJ \ge 0$. The additional term increases a possible value of the gained energy.  This increase has a pure informational character since  $\infJ =\info_{\GhI} (\GsI )$
is the value of information about such microscopic initial conditions leading to the demanded gain of energy.

It is instructive to use the inequality (\ref{WF2}) in a situation when an observer repeats the experiment in which the initial observed indicator is always $\ywI$ and the probability of occurrence of a concrete initial state of internal degrees of freedom (microstate), $\stI$, corresponds to the state of thermal equilibrium, $p_{eq}(\ha_{int}(\stI ,\ywI)) $. If $\GsI\neq\GhI$ the resulting value of $\yw$ is not always $\ywF$ and various results of $\Delta\E=\W_{u}(\stI )$ can be expected. The observed averaged quantity, 
$$\langle{\W_{u}}\rangle = \sum_{\stI\in\Gh} p_{eq}(\stI, \ywI )\W_{u}(\stI ), 
$$
where $\W_{u}(\stI )=\E (\yw (\stI ))-\E (\ywI)$ with $\yw (\stI )$ being the final value of $\yw$ if the initial $\st$ is $\stI$ (if $\stI\in\GsI$ then $\yw (\stI )=\ywF$, see Fig.~(\ref{fig:2})). Other averaged values  are defined similarly.

The use of Eq.~(\ref{Hinf}) for determining $\infJ=\info_{\GhI} (\GsI )$ and averaging Eq.~(\ref{WF2}) gives the averaged form of this inequality, namely 
\begin{equation}\label{GSL}
\langle{\W_{u}}\rangle\le \langle \Delta F\rangle +k_BT I,
\end{equation}
where $I =\langle{\infJ}\rangle$ is the mutual information. The inequality  Eq.~(\ref{GSL}) is called the generalized second law of thermodynamics \cite{SagUed2009,SagUed2010,SagUed2013}.
It is worth stressing that this concrete form is valid in the case when the averaging is done over the initial distribution corresponding to the thermal equilibrium. Its information form, Eq.~(\ref{WF2}), is valid for a single adiabatic  process on an arbitrary length scale with an arbitrary initial conditions (e.g. that  representing a highly nonequilibrium macroscopic state).

\section{Concluding discussion}

The statistical interpretation of  the second law of thermodynamics is the most natural and logical explanation of why reversible microscopic behavior can manifest as irreversible at macroscopic scales. It is only a statistical reason why a drop of ink put into a bottle with water will always smear over the water, and why the  dissolved ink does not return into the initial drop formation. Namely the macrostate with a higher Boltzmann entropy is overwhelmingly larger then that with a lower one (see Fig.~\ref{fig:4}a). Hence the microstate wanders  into this huge set in overwhelmingly many cases \cite{Penrose2005}. 

I argue that the statistical  way of analyzing macroscopic processes at the microscopic level is nothing but an expression of information ideas. The set of possible microstates $\Gh$ (e.g. Boltzmann's macrostate) is information about the actual microstate $\xx$: it belongs into $\Gh$. Entropy is information about the actual microstate, $\SI \sim \info (\xx )$. The crucial idea is that the \emph{value} of information (as well as value of other physical quantities) can be defined only with respect to referential information. Contrary to the majority of physical quantities that can be defined with respect to a firmly given referential element, the referential sets $\Gh$ are usually defined by actual values of some physical quantities (observable indicators), $\yw$, that vary in time. Moreover, the referential set changes (diminishes) whenever the observer (a human, an experimental device, robot, etc.) gains new information about the studied system. The information reference frame (IRF) formed by referential sets $\Gh (t)$ thus resembles more referential frames of the general theory of relativity firmy connected with dynamics of physical fields \cite{Rovelli2004}.

The essential element of the information form of the second law of thermodynamics is the value of information gained at time $t$ that concerns the coarse-grained state of the system at another time $t'$. Let the observer know only the actual coarse-grained state at $t$, say the current values of some physical quantities $\yw$. If she gets the message that this state will be/was $\yw '$ at $t'$ the value of this message, $J$, depends on $\yw$ and $\yw '$  ($t,t'$ are given). This message reveals the process between $t$ and $t'$ if the situation at $t$ is known, i.e. we can write $J=J (\yw  ,\yw \to\yw ')\ge 0$. The second law of thermodynamics for  adiabatic processes, Eq.~(\ref{entropy}), then can be written as 
\begin{equation}\label{ISL}
k_B^{-1}(\SI (t')-\SI (t))= J (\yw',\yw '\to\yw ) -J (\yw ,\yw \to\yw ').
\end{equation}
In the macroscopic limit, $|\Delta\SI|/k_B\to\infty$,
the value $J(\yw ,\yw\to\yw ')\to \infty$ if $S(t')<S(t)$, i.e. $P(\yw,\yw\to\yw ')\to 0$ due to Eq.~(\ref{PPP}). Hence the observer who knows that the observable indicator is $\yw$ at $t$ must conclude that the observable indicator at $t'$ cannot  be  $\yw '$. 

It looks paradoxically  since we must get either $P(\yw ,\yw\to\yw ')\to 0$ or $P(\yw ' ,\yw '\to\yw )\to 0$ if $\Delta\SI \neq 0$ in the macroscopic limit, hence the occurrence of any couple of coarse-grained states $\yw$, $\yw '$ differing by a nonzero entropy appears as impossible. The explanation of this evident nonsensical result 
consists in the fact that the past and future are not symmetric: the observer can verify the situation at $t'$ if $t'>t$. It implies that if $P(\yw ,\yw\to\yw ')\to 0$ and $t'>t$ the observer \emph{cannot} detect $\yw '$ at $t'$. Eq.~(\ref{ISL}) thus implies that the adiabatic evolution of a macroscopic system cannot be connected with a macroscopic decrease of entropy.  

If $t'<t$ the fact that $P(\yw ,\yw\to\yw ')\to 0$ does not make a controversy since the observer who detects $\yw$ at $t$ then cannot  verify the situation at $t'$ (we cannot move  into the past). 
If there is, however, information that $\yw$ \emph{was} $\yw '$ at $t'$ available at $t$, this information must occur in form of a record (e.g. a change in some brain cells) that exists  at $t$. The observer at $t$ thus has information about the system that forms the dynamic referential frame IRF' in which $\Delta\SI '=0$ and Eq.~(\ref{ISL}) implies that $P'(\yw ',\yw '\to\yw)=P'(\yw ',\yw '\to\yw)$ may have an arbitrary value. The information form of the second law thus inheres  an important referential  asymmetry between past and future though no such asymmetry occurs in  microscopic physics \cite{Holecek2022}. 

The important question is why  the change of the referential frame, IRF~$\to$~IRF', "automatically" happens whenever a record about the past appears. The explanation consists in the fact that the information reference frame is primarily defined by \emph{information}. A collection of subsets $\Gh (t)\subset\GA$ is only its mathematical expression. A record is a physical event that can be interpreted as information about a past moment. If so, any  physical consideration must take into account  this information. 

An analogy with spatial reference frames is instructive here.
Namely  the reference frame in space is primarily done by a collection of referential bodies (an analogy of attainable information about a physical system). These bodies define the mathematical structure ${\cal R}^3$ denoting coordinates of individual spatial points (an analogy with the sets $\Gh (t)$). Whenever the original configuration of the referential bodies changes  the same physical point is described by other coordinates. Analogically, whenever information about the system changes the same physical situation is described via different sets $\Gh (t)$.

The information form of the second law of thermodynamics for adiabatic systems, $\Delta\SI =k_B\Delta\infJ$, is valid at arbitrary length scales.  In probabilistic interpretation, it can be formulated by Eq.~(\ref{pp}) that expresses some stochastic character of the gained results  whenever information about the system is incomplete (i.e. when the sets $\Gh (t)$ does not include only the actual microstate $\xx (t)$). 

Let us denote $\xx^R$ the time reversed state of $\xx$ (e.g. the state in which velocities of all particles have an opposite sign) and $\yw^R$ be the time reversed state of $\yw$ with $\Gh (\yw^R)=\{\xx^R, \xx\in\Gh (\yw )\}$. 
If $\pout$ is interpreted as the probability  $P_{rev}$ of realizing the reversed process $\ywF^R\to\ywI^R$ then Eq.~(\ref{PPP}) has  the form 
\begin{equation}\label{FT}
\frac{P_{rev}}{P}=\e^{-\Sigma},
\end{equation}  
where $\Sigma =(k_BT)^{-1}\W_{dis}$ in $\ha$-representation. We thus get the information form of  fluctuation theorem  \cite{Sev2008,SeiHafJar2021}. Its content is, however, somewhat different from the standard fluctuation theorems. Namely it concerns the adiabatic processes only and the probabilities are defined with respect to the used IRF.

There are many conceptual questions concerning the presented approach. For example, the implementation of 'information' as a full valuable physical quantity means a change of viewpoint concerning the meaning of some physical quantities. Namely when  some physical event (structure, configuration) is interpreted as information about the studied system we immediately  get different  values of quantities like entropy or free energy since the information reference frame changes. This effect does not seem to play a role in macroscopic physics where the information reference frame is usualy fixed (given by a typical macroscopic observer).  It may be important, however, in microphysics: the role of 'observer' (whatever it means) is nontrivial here and transfromations between various reference frames might play an essential role. It corresponds to recent discoveries in microscopic statistical physics underlaying the crucial role of information, for example, in energy conversion  \cite{ParHorSag,HorSagPar2013,PanYunTluPak2018}.

\section*{Appendix}

The equality Eq.~(\ref{proof2}) holds for all sufficiently small $\tau >0$. Since $\lim_{\tau\to 0} \Delta\infJ =0$  we denote  $H(x_1,x_2,\ldots )=\F( \F_{-1}(x_1)+ \F_{-1}(x_2)\ldots )$ and study  the condition $H(x_1+y,x_2+y,\ldots )=H(x_1,x_2,\ldots )+y$ in the limit $y\to 0$. The derivation of this condition by $y$ gives $\phi (\sum \F_{-1}(x_i))=\sum \phi ( \F_{-1}(x_i))$ in the limit $y\to 0$, where  
 $\phi (Z)\equiv (\F'(Z))^{-1}$. It implies that $\phi (Z)=\kappa Z$, where $\kappa$ is a constant. After integration we get $\F (Z)=-\del\ln Z+\kappa_0$, where $\del=-\kappa^{-1} > 0$ ($\F$ is a decreasing function) which implies $f(\xa)=\e^{-\del^{-1}\QQ (\xa ) }\e^{\del^{-1}\kappa_0}$.    
Hence $\info_{\Gh}(\Om ) = -\del\ln\sum_{\xa\in\Om}\e^{-\del^{-1}k_B^{-1}\SI_\Gh(\xa)}$, where $\del$ cannot depend on $\Gh$ to fulfill Eq.~(\ref{It}) for $\GhI \neq\GhF$.

\section*{Acknowledgement}

The author is indebted to J\'{a}n Min\'{a}r for his help and constructive discussions, and to Philipp  Strasberg for inspirational and critical comments concerning the preparation of the manuscript. The work is supported by the New Technologies Research Center of the West Bohemia University in Pilsen.

\end{document}